\documentclass[aps]{revtex4}
\usepackage{color}
\usepackage{graphicx}
\usepackage{eurosym}
\usepackage{graphicx}
\usepackage{amsmath,amssymb}
\usepackage{color}
\usepackage[outdir=./]{epstopdf}
\usepackage{amsmath}
\usepackage{amsfonts}
\usepackage{amssymb}

\def\bc{\begin{center}}
\def\ec{\end{center}}

\def\beq{\begin{eqnarray}}
\def\eeq{\end{eqnarray}}

\def\bc{\begin{center}}
\def\ec{\end{center}}

\def\beq{\begin{eqnarray}}
\def\eeq{\end{eqnarray}}

\begin{document}

\title{Strain induced quantum Hall phenomena of excitons in graphene}
\author{Oleg L. Berman$^{1}$, Roman Ya. Kezerashvili$^{1}$, Yurii E. Lozovik$%
^{2}$, and Klaus G. Ziegler$^{3}$}
\affiliation{\mbox{$^{1}$Physics Department, New York City College
of Technology, The City University of New York,} \\
Brooklyn, NY 11201, USA \\
\mbox{$^{2}$Institute of Spectroscopy, Russian Academy of
Sciences,} \\
142190 Troitsk, Moscow, Russia \\
\mbox{$^{3}$Institut f\"ur Physik, Universit\"at Augsburg,\\
D-86135 Augsburg, Germany }}
\date{\today }

\begin{abstract}
We predict quantum Hall phenomena for neutral composite bosons - excitons
such as the Integer and Fractional Quantum Hall effects, and the state of
composite fermions at $\nu = 1/2$ in a mono and double layer of gapped
graphene under strain induced gauge pseudomagnetic field. When electrons and
holes are excited only in one valley of the honeycomb lattice of gapped
graphene, the strain induced pseudomagnetic field acts on electrons and
holes the same way. The latter leads to the formation of single
pseudomagnetoexcitons (PMEs) whose properties are cardinally different from
those of magnetoexcitons. Wave functions and the energy spectrum of direct
and indirect PMEs in a mono and double layer of gapped graphene are
obtained. Experimental observations of the quantum Hall phenomena for the
PMEs are discussed.
\end{abstract}

\maketitle

In this study we focus on the influence of strain on the properties of
excitons in mono and double layers of gapped graphene. It was predicted that
an in-plane distortion of the graphene lattice due to non-uniform strain is
equivalent to creation of large, nearly uniform pseudomagnetic fields,
acting on electrons, which lead to the formation of Landau levels and zero
magnetic field Quantum Hall effect (QHE) for electrons~\cite%
{Geim_2010,Guinea_Geim_2010,Amorim,Vozmediano_2008}. Note that the Quantum
Hall phenomena  in graphene and graphene based structures in a high magnetic
field attracted the great interest~\cite{Jain2019,Halperin2019,Dean2019}.
Landau quantization of the electronic spectrum for highly strained
nanobubbles was experimentally observed, and pseudomagnetic fields in excess
of $300 \ \mathrm{T}$ have been measured \cite{Guinea_Levy_2010}.
A two-dimensional electron gas (2DEG) in graphene in a high strain-induced
pseudomagnetic field attracted strong interest very recently~\cite{Shalom}.

Since the strain induced pseudomagnetic field in contrast to a magnetic
field acts on the  different electric charges the same way, it leads to
novel effects for excitons in pseudomagnetic fields as predicted below.
These effects are (1) the completely discrete energy spectrum of excitons in
a pseudomagnetic field in contrast to excitons in magnetic field,
characterized by the continues energy dependence on magnetic momentum on
each Landau level~\cite{Lerner,Ruvinsky}; (2) rich Quantum Hall phenomena
for excitons in a pseudomagnetic field which are absent for excitons in a
magnetic field. In this Letter we calculate the Landau levels of direct and
indirect excitons in mono and double layers of gapped graphene,
respectively, in the presence of a strain induced pseudomagnetic field. Such
excitons we refer as pseudomagnetic excitons (PMEs). We predict the
existence of the Integer and Fractional Quantum Hall effects and the state
of composite fermions at $\nu = 1/2$ for the PMEs. We propose also the
experimental methods to observe these phenomena.

\begin{figure}[t]
\caption{(Color online) Conceptual picture of the quantum Hall effect of
PMEs in graphene. The QHE PME flow, shown by the blue arrow, is generated by
the strain induced pseudomagnetic field $\mathbf{B}$, normal to the graphene
layer, and a pseudoelectric field $\mathbf{E}_{\mathrm{pseudo}}$ or $-%
\protect\nabla T$, or $-\protect\nabla \protect\mu$, directed along the
graphene layer. The QHE PME flow in the graphene layer is directed normal to
$\mathbf{E}_{\mathrm{pseudo}}$ or $-\protect\nabla T$, or $-\protect\nabla
\protect\mu$. The strain is created by the normal forces applied at two
opposite boundaries and their magnitude is gradually decreased, which is
indicated by the length of the plotted arrows.}
\label{Fig1}
\centering
\includegraphics[width=3.5in]{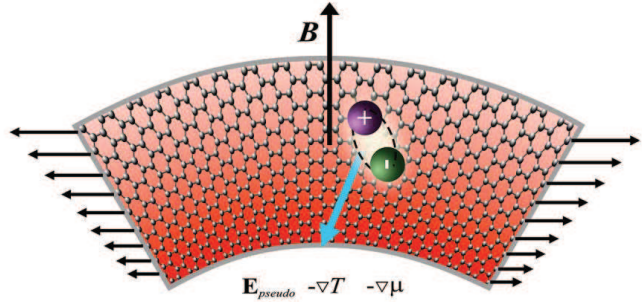} %
\end{figure}

The conceptual picture of the quantum Hall effect of PMEs is presented in
Fig.~\ref{Fig1}. The suggested bending geometry, when the graphene rectangle
is bent into a circular arc, would generate a uniform pseudomagnetic field~%
\cite{Guinea_Geim_2010,another}. When graphene is under the strain, shear
strain induces a pseudomagnetic field, while the dilatation gives rise to an
effective scalar potential which results in the pseudoelectric field, acting
on the charge carriers independently of the sign of charge, contrary to
ordinary electric field~\cite{Guinea_Geim_2010}. The pseudoelectric field $%
\mathbf{E}_{\mathrm{pseudo}}$ can be chosen to be normal to the
pseudomagnetic field. The other option is to use laser illumination on the
edge of samples, which creates gradients of the temperature $T$ and/or
chemical potential $\mu $, which drive electrons ($e$) and holes ($h$) in
the same direction. This can trigger the flow of PMEs without breaking the
bound states of $e$ and $h$. Below we show that $\mathbf{E}_{\mathrm{pseudo}}
$ or $\nabla \mu$, or $\nabla T$ together with the pseudomagnetic field
initiates the Quantum Hall effect for the PMEs. Note that contrary to PMEs,
the electric field normal to the magnetic field does not lead to any flow of
magnetoexcitons (MEs), but only induces the dipole moments of MEs. Lets us
mention that $\nabla \mu$ and $\nabla T$ induce flows of MEs in the
directions of the gradients, without creating Hall flows in contrast to PMEs.

As a first step, we consider the existence of excitons in a strain-induced
pseudomagnetic field and obtain their eigenfunctions and eigenenergies.
Starting from the effective Dirac Hamiltonian for a single valley for gapped
graphene in the presence of a strain field with components $A_{x}$ and $A_{y}
$~\cite{Amorim}, we consider here the Dirac equation for a pair of an
electron at position $\mathbf{r}_{1}$ and a hole at position $\mathbf{r}_{2}$
as
\begin{equation}
\left( H_{0}+\ V\left( \left\vert \mathbf{r}_{1}-\mathbf{r}_{2}\right\vert
\right) \right) \Psi =\mathcal{E}\Psi ,\ \ \   \label{General}
\end{equation}%
with
\begin{equation}
H_{0}=v_{F}\sum_{j=1}^{2}\left(
\begin{array}{cc}
2 \Delta/v_{F} & i\hbar \partial _{x_{j}}+A_{x}(\mathbf{r}_{j})+\hbar
\partial _{y_{j}}-iA_{y}(\mathbf{r}_{j}) \\
i\hbar \partial _{x_{j}}+A_{x}(\mathbf{r}_{j})-\hbar \partial
_{y_{j}}+iA_{y}(\mathbf{r}_{j}) & -2 \Delta/v_{F}%
\end{array}%
\right) \text{, \ }\Psi =\left(
\begin{array}{c}
\psi _{1}(\mathbf{r}_{1},\mathbf{r}_{2}) \\
\psi _{2}(\mathbf{r}_{1},\mathbf{r}_{2})%
\end{array}%
\right) .  \label{dirac00}
\end{equation}
In Eq. (\ref{dirac00}) $H_{0}$ is the Hamiltonian of the non-interacting
electron and hole in gapped graphene 
in the presence of the strain field, where $v_{F}$ is the Fermi velocity,
while the wave function $\Psi $\ can be understood as a two-component
spinor.
The potential energy of electron-hole attraction $V\left( \left\vert \mathbf{%
r}_{1}-\mathbf{r}_{2}\right\vert \right) $ in Eq. (\ref{General}) can be
described either by the Coulomb potential or, taking into account
polarization of a 2D material, by the Rytova-Keldysh (RK) potential~\cite%
{Rytova,Keldysh}. In stark contrast to the vector potential of the
electromagnetic field, the strain induced effective vector potentials $%
\mathbf{A}(\mathbf{r}_{1})$ and $\mathbf{A}(\mathbf{r}_{2})$, acting on an
electron and a hole, forming PME, are not coupled to the charges of the
particles and have the same sign in the Hamiltonian~(\ref{dirac00}), and
these potentials act on an electron and a hole the same way, and the
quasi-Lorentz force due to strain induced pseudomagnetic field can be
exerted on a moving PME. Note that the units of $\mathbf{A}(\mathbf{r}%
_{1(2)})$ and $\mathbf{B}$ in SI system are $\mathrm{kg\times m/s}$ and $%
\mathrm{kg/s}$, correspondingly, while the units of $\mathbf{B}/e$, where $e$
is an electron charge, are Teslas. As the first step, we consider
non-interacting electron and hole in the stain-induced field and find their
eigenfunctions and eigenenergies: $H_{0}\Psi _{0}=\mathcal{E}_{0}\Psi _{0}$.
In result, we obtain for  $\psi _{0s}(\mathbf{r}_{1},\mathbf{r}_{2})=\psi
_{s,1}(\mathbf{r}_{1})\psi _{s,2}(\mathbf{r}_{2})$ and $\mathcal{E}_{0}=%
\mathcal{E}_{1}+\mathcal{E}_{2}$ for the two-component strain field $\mathbf{%
A}=(A_{x},A_{y})$ in the case of a uniform pseudomagnetic field $%
B_{z}=(\nabla \times \mathbf{A})_{z}$ (see Supplementary Material S1)
\begin{eqnarray}
v_{F}^{2}[i\hbar \nabla _{\mathbf{r}_{j}}+\mathbf{A}(\mathbf{r}%
_{j})]^{2}\psi _{1,j}=E_{j}\psi _{1,j}\ ,\ \ \ E_{j}=\mathcal{E}%
_{j}^{2}+\hbar v_{F}^{2}B_{z}-4\Delta ^{2}\ .  \label{schr-eq}
\end{eqnarray}

The energy $\mathcal{E}_{0}$ is related to the eigenvalue $E_{1}$ ($E_{2}$)
of the electron (hole) from Eq. (\ref{schr-eq}) via
\begin{equation}
\mathcal{E}_{0}=\sum_{j=1}^{2}\sqrt{E_{j}-\hbar v_{F}^{2}B_{z}+4\Delta ^{2}}%
\ .  \label{Dirac_spectrum}
\end{equation}%
By considering the non-interacting electron and hole the homogeneous $B_{z}$
field \cite{Landau} Eq. (\ref{Dirac_spectrum}) becomes
\begin{equation}
\mathcal{E}_{0n_{1},n_{2}}=\sqrt{2\hbar v_{F}^{2}B_{z}n_{1}+4\Delta ^{2}}+%
\sqrt{2\hbar v_{F}^{2}B_{z}n_{2}+4\Delta ^{2}},  \label{Landau_spectrum}
\end{equation}
where $n_{1},$ $n_{2}=0,1,2,...$ are the quantum numbers, corresponding to
the Landau levels for the electron and the hole.

In monolayer gapped graphene the effective electron and hole masses are
defined as $m_{0}=\Delta /v_{F}^{2}$~\cite{Pedersen} and two Schr\"{o}dinger
equations Eq. (\ref{schr-eq}), one for the electron and one for the hole,
can be combined as
\begin{equation}
\sum_{j=1}^{2}\left[ i\hbar \nabla _{\mathbf{r}_{j}}+\mathbf{A}(\mathbf{r}%
_{j})\right] ^{2}\psi _{1,1}(\mathbf{r}_{1})\psi _{1,2}(\mathbf{r}_{2})=%
\frac{1}{v_{F}^{2}}\left( E_{1}+E_{2}\right) \psi _{1,1}(\mathbf{r}_{1})\psi
_{1,2}(\mathbf{r}_{2})\ .  \label{schroedinger2}
\end{equation}
Let us mention that one can consider $e$ and $h$ in different layers, which
form indirect~\cite{Ruvinsky} PMEs in a graphene double layer. In this case,
the application of different doping to the two graphene monolayers can lead
to the formation of two different gaps in these graphene monolayers. In this
system, the effective masses $m_{j}=\Delta _{j}/v_{F}^{2},$ of $e$ and $h$
can be unequal: $m_{1}\neq m_{2}$ (see Supplementary Material).

The Schr\"{o}dinger equations for a ME in a magnetic field and for the PME
in a pseudomagnetic field are invariant with respect to the translation and
the gauge transformations. The invariance for a magnetic field results in
the conservation of the operator of the magnetic momentum of the ME $\hat{%
\widetilde{\mathbf{P}}}=-i\hbar \nabla _{\mathbf{r}_{1}}-i\hbar \nabla _{%
\mathbf{r}_{2}}-\frac{e\mathbf{B}_{0}\times \left( \mathbf{r}_{1}-\mathbf{r}%
_{2}\right) }{2}$~\cite{GD,Lerner,Ruvinsky} and in the conservation of the
operator of pseudomagnetic momentum $\hat{\mathbf{P}}=-i\hbar \nabla _{%
\mathbf{r}_{1}}-i\hbar \nabla _{\mathbf{r}_{2}}-\frac{\mathbf{B}\times
\left( \mathbf{r}_{1}+\mathbf{r}_{2}\right) }{2}$ of the PME in the
pseudomagnetic field. Since the operators $\hat{\widetilde{\mathbf{P}}}$ and
the Hamiltonian of a ME, as well as $\hat{\mathbf{P}}$ and the Hamiltonian
of a PME commute, these pairs of operators have the same eigenfunctions,
respectively.  The eigenfunctions are different for ME and PME because the
pseudomagnetic field acts the same way on $e$ and $h$. The latter leads to
the fact that the spectrum of a PME in a high strain-induced pseudomagnetic
field is fully discrete (see Supplementary Material), in contrast to the
spectrum of a ME in a high magnetic field which depends continuously on the
magnetic momentum~\cite{GD,Lerner}. In result, the collective properties of
a 2D many-PME system contrary to a 2D many-ME system occur to be similar to
a 2D electron system in a high magnetic field.

Expressing the l.h.s. of Eq.~(\ref{schroedinger2}) in terms of the operator $%
\hat{\mathbf{P}}$ for a PME, after lengthly calculations presented in
Supplementary Material S2 and S3 for more general case for unequal $%
m_{1}\neq m_{2}$ masses, we obtain
\begin{equation}
H=\frac{1}{4m_{0}}\left[ \hat{\mathbf{P}}^{2}- \frac{1}{m_{0}^{2}}\left( 2
\hbar m_{0} \nabla _{\mathbf{r}}-i\mathbf{S}(\mathbf{r})\right) ^{2}\right]
,\ \ \mathbf{S}(\mathbf{r})=\frac{m_{0}}{2}\mathbf{B}\times \mathbf{r}.
\label{HamDP1}
\end{equation}
The corresponding eigenfunctions and eigenenergies of Hamiltonian (\ref%
{HamDP1}) for the non-interacting electron and hole with different effective
masses in the strain induced pseudomagnetic field are obtained 
in Supplementary Material S3. Using the Hamiltonian (\ref{HamDP1}) for the
non-interacting electron and hole one can write the Hamiltonian $\hat{H}%
_{eh} $ for the PME \ in the Schr\"{o}dinger picture as
\begin{equation}
\hat{H}_{eh}=\hat{H}+V\left( \left\vert \mathbf{r}_{1}-\mathbf{r}%
_{2}\right\vert \right) .  \label{Hamehv}
\end{equation}%
Assuming that contribution of $V\left( \left\vert \mathbf{r}_{1}-\mathbf{r}%
_{2}\right\vert \right) $ to the energy of the PME (the binding energy of
the PME) is small compared with the difference between the eigenvalues of $%
\hat{H}$, we start with the zeroth order of the perturbation theory with
respect to $V\left( \left\vert \mathbf{r}_{1}-\mathbf{r}_{2}\right\vert
\right) $ and find the eigenvalues and eigenfunctions for (\ref{HamDP1}).
Later we will take into account the electron-hole interaction $V\left(
\left\vert \mathbf{r}_{1}-\mathbf{r}_{2}\right\vert \right) $ within the
first order of the perturbation theory.

The wavefunction of the electron-hole pair in the strain induced
pseudomagnetic field, neglecting the electron-hole attraction, can be
written as $\Psi _{n,m,\tilde{n},\tilde{m}}(\mathbf{R},\mathbf{r})=\psi
_{n,m}^{(0)}(\mathbf{R})\tilde{\varphi}_{\tilde{n},\tilde{m}}^{(0)}(\mathbf{r%
})$, where $\psi _{n,m}^{(0)}(\mathbf{R})$ and $\tilde{\varphi}_{\tilde{n},%
\tilde{m}}^{(0)}(\mathbf{r})$ are the wavefunctions for a free particle in
the pseudomagnetic field $2\mathbf{B}$ and $\mathbf{\tilde{B}}=\mathbf{B}/2$%
, respectively, in the cylindrical gauge \cite{Landau,Lerner,Ruvinsky}. The
functions $\tilde{\varphi}_{\tilde{n},\tilde{m}}^{(0)}(\mathbf{r})$ are
defined in Supplementary Material S3. Using the pseudomagnetic fields $B_{z}
$ and ${\tilde{B}}_{z}=B_{z}/2$ in the Schr\"{o}dinger equation with
center-of-mass and relative coordinates (cf. Supplementary Material), and $%
m_{0}=\Delta /v_{F}^{2}$ for electrons and holes masses, we obtain  a simple
elegant expression:
\begin{equation}
\mathcal{E}_{0n,\tilde{n}}=\sqrt{4\hbar v_{F}^{2}B_{z}n+16\Delta ^{2}}+\sqrt{
\hbar v_{F}^{2}B_{z}{\tilde{n}}+\Delta ^{2}},  \label{calEn}
\end{equation}%
where $n=0,1,2,\ldots $ and $\tilde{n}=\min (n_{1},n_{2})$ are the quantum
numbers for the motion of the center-of-mass and the relative motion of an
electron and a hole \cite{Ruvinsky} in the PME, respectively. In the case of
the double layer when  it is also possible that $m_{1}\neq m_{2}$, Eq. (\ref%
{calEn}) can be written in the form shown in Supplementary Material S3.
Therefore, Eq.~(\ref{calEn}) presents the quantized eigenenergy of the
non-interacting electron and hole in the strain induced pseudomagnetic
field. Thus, the energy levels of a non-interacting two-dimensional
electron-hole system in a pseudomagnetic field are quantized into a discrete
set of Landau levels with degeneracy proportional to the area of the system~%
\cite{QHE}.

Now let us find the energies of a PME in a mono and double layer of gapped
graphene in the presence of the electron-hole attraction. The attractive
electron-hole interaction is treated in the framework of the perturbation
theory. Neglecting the transitions between different Landau levels, the
first order perturbation with respect to electron-hole attraction results in
the following expression for the energy $E_{\tilde{n},\tilde{m}}$ of a PME:
\begin{equation}
E_{\tilde{n},\tilde{m}}^{\prime}=\left\langle \Psi _{n,m,\tilde{n},\tilde{m}%
}(\mathbf{R},\mathbf{r})\left\vert V(r)\right\vert \Psi _{n,m,\tilde{n},%
\tilde{m}}(\mathbf{R},\mathbf{r})\right\rangle =\left\langle \tilde{\varphi}%
_{\tilde{n},\tilde{m}}^{(0)}(\mathbf{r})\left\vert V(r)\right\vert \tilde{%
\varphi}_{\tilde{n},\tilde{m}}^{(0)}(\mathbf{r})\right\rangle .
\label{Enmag}
\end{equation}
One can calculate the energies of a direct and an indirect PME using the
Coulomb and Rytova-Keldysh potentials. By substituting these potentials into
Eq.~(\ref{Enmag}), and using the wavefunctions for the corresponding state
given in Supplementary Material S4, we obtain the analytical expressions for
the eigenenergies $E_{0,0}$, $E_{0,1}$ and $E_{1,0}$. The total energy $E_{n,%
\tilde{n},\tilde{m}}^{(tot)}$ of a direct PME in gapped graphene monolayer,
taking into account electron-hole attraction, is
\begin{eqnarray}
\mathcal{E}_{n,\tilde{n},\tilde{m}}=\mathcal{E}_{0n,\tilde{n}}+E_{\tilde{n},%
\tilde{m}}^{\prime},
\end{eqnarray}
where $\mathcal{E}_{0n,\tilde{n}}$ is given by Eq.~(\ref{calEn}), and
analytical expressions for the energy $E_{\tilde{n},\tilde{m}}^{\prime }$ of
direct PMEs for the Coulomb and the RK potentials, are given in
Supplementary Material S5. The energy of an indirect PME in a double layer
of gapped graphene can be obtained by substituting the Coulomb or RK
potentials into Eq. (\ref{Enmag}). In the case of the Coulomb potential the
corresponding matrix elements can be evaluated analytically and the results
are given in Supplementary Material S6. However, in the case of the
Rytova-Keldysh potential the energy of an indirect PME in a double layer
system could be found only numerically.

\begin{figure}[t]
\caption{(Color online) The dependence of energies $E_{\tilde{n},\tilde{m}%
}^{\prime }$ of indirect PMEs on the separation $D$ between gapped graphene
layers. Calculations performed for the value of magnetic length $l$ that
corresponds to $B/e=50\ \mathrm{T}$. }
\label{Fig2}
\centering
\includegraphics[width=5.0in]{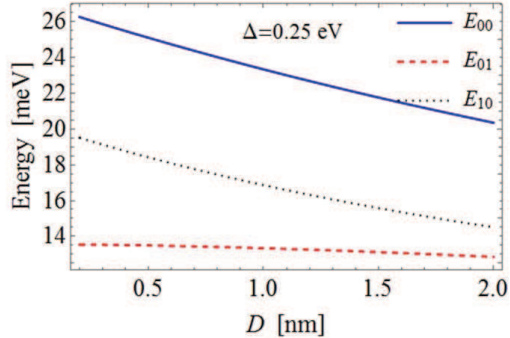} %
\vspace{-1cm}
\end{figure}

In our calculations the uniform pseudomagnetic field $B$, acting on an
electron or a hole, is related to the strain as $B=\frac{8\hbar \beta c}{a}$
\cite{Geim_2010}, where $a=2.566\ {\mathring{A}}$ is the lattice constant~%
\cite{Lukose}, and the parameter $\beta \approx 2$ and the constant $c$ are
defined in Ref.~\cite{Geim_2010}. It can be seen from Fig.~\ref{Fig2} that
the energies $E_{\tilde{n},\tilde{m}}^{^{\prime }}$ of PME's are decreasing
with the increase of the separation between graphene layers.
Interesting enough, the comparison of results for the binding energies of
the direct and indirect PME energies, calculated using the RK and Coulomb
potential, for the parameters used are very close and almost the same. Also,
the magnitudes of PME energies, calculated using the RK potential are a
little bit smaller than ones, calculated using the Coulomb potential,
because the RK potential implies the screening effects. The energies $E_{%
\tilde{n},\tilde{m}}$ of indirect PMEs as a function of the separation $D$
between gapped graphene layers are
presented in ~Fig.~\ref{Fig2}. The 3D plot of the dependence of $E_{\tilde{n}%
, \tilde{m}}^{\prime }$ on the separation $D$ and pseudomagnetic field $B$
is presented in Supplementary Material S6. It can be seen, the energies of
indirect PMEs decrease with the increase of $D$ and increase with the
increase of $B$.

A strain induced pseudomagnetic field leads to quantization of the PME
spectrum which is discrete. Scattering effects due to random potentials
and/or interaction between the PMEs would lead to a broadening of the Landau
levels. The resulting multiband structure can be characterized by the Chern
numbers of these PME bands, which can be observed in the transport
properties. The strain-induced  pseudoelectric field causes a flow of the
PMEs inside the graphene layers. The strain-induced pseudomagnetic and
pseudoelectric field or existence of $\nabla \mu$, or $\nabla T$ give rise
to the Hall effect for the PMEs, whose Hall conductivity is quantized
according to the Chern numbers of the multiband structure. This can also be
formulated in terms of an effective Ginzburg-Landau approach by coupling an
additional statistical Chern-Simons gauge field to the bosonic PMEs \cite%
{zhang89,fradkin_book}, where the resulting Hall conductivity is related to
the Chern-Simons constant. Thus, analogously to the standard Integer Quantum
Hall effect (IQHE) for the 2D electron gas (2DEG) in a magnetic field~\cite%
{QHE}, we obtain for the system of the PMEs in the presence of impurities
the occurrence of the set of plateaus in the Hall resistivity $\varrho_{xy}$
and conductivity $\sigma_{xy}$ with quantized values: $\varrho_{xy} =
-1/\sigma_{xy} = h/n$ ($n=1,2,3, \ldots$ is the Landau level for the motion
of the center-of-mass of the PME). In the region of the plateau we have $%
\varrho_{xx} = \sigma_{xx} = 0$.

The degeneracy $d$ of the Landau levels $n$
is given by $d = 2BS/\Phi_{0}$, where $\Phi_{0} = h/2$ is the quantum of
pseudomagnetic flux, $S$ is the area of the system, and the factor $2$
appears due to the same action of the pseudomagnetic field an electron and a
hole. Thus, one can control the filling factor of the Landau level $\nu = N/d
$ ($N$ is the number of the PMEs) either by changing the strain, inducing
the pseudomagnetic field, or by laser pumping changing the number $N$ of the
PMEs. So one can observe not only the IQHE but also the Fractional Quantum
Hall effect (FQHE) for PMEs. For example, to observe the FQHE at the filling
factor $\nu = 1/3$, it occurs that one needs the pseudomagnetic field $B$
corresponding to four (but not three as for the 2DEG in a magnetic field)
quanta of pseudomagnetic flux accounting for one bosonic PME. In this case,
a composite fermion can be formed via attachment of one pseudomagnetic flux
quantum to one PME, and these composite fermions with three remaining
pseudomagnetic flux quanta form the FQHE state at $\nu = 1/3$ analogously to
the FQHE for the 2DEG in a magnetic field~\cite{Jain}. Note that we also can
achieve the state of the PME system analogous to the state of composite
fermions at the filling factor $\nu = 1/2$~\cite{HLR,EM}, when the filling
factor of PMEs $\nu = 1/3$. Really the latter corresponds to three
pseudomagnetic flux quanta on one PME. When one pseudomagnetic flux quantum
is attached to each boson, PME, one obtains the system of the composite
fermions at the filling factor $\nu = 1/2$ (experimentally observable
composite fermions, analogously to electrons with two attached magnetic flux
quanta \cite{HLR}, correspond to the PMEs with three attached pseudomagnetic
flux quanta). Thus, for the neutral PMEs (bosons) one can observe the IQHE,
FQHE, and $\nu = 1/2$ phenomena similar to the ones for charged electrons,
forming 2DEG in the high magnetic field~\cite{Jain}.

It should be mentioned that by breaking the isotropy in the honeycomb
lattice strain creates a pseudomagnetic field for each of the two Dirac
cones, but it does not break the time-reversal symmetry. By reversing the
time the momentum changes the sign, and two valleys are switched. Therefore,
a strain induced pseudomagnetic field induces the exciton Hall valley flows
in opposite directions in two valleys of graphene 
and the total Hall exciton flow is absent due to the time-reversal symmetry,
if the excitons in both valleys are excited. However, the Hamiltonian~(\ref%
{schroedinger2}) for an electron-hole pair corresponds to creation of
excitons in only one valley of gapped graphene by the circularly polarized
pumping beam~\cite{Xiao,Mak2013}. This circularly polarized pumping beam
breaks the time-reversal symmetry, and a pseudomagnetic field induces the
Hall exciton flows in a monolayer or the Hall electron and hole currents in
two layers, forming a double layer, in only one valley of gapped graphene.
Thus, Hall valley flows of direct and indirect excitons similar to Hall
currents of charged particles can be excited in a mono or double layer of
the gapped graphene, respectively. These exciton Hall valley flows, excited
in two valleys of graphene 
are characterized by opposite directions, corresponding to the excitons in
each valley. However, we consider the scenario, when the charge carriers are
excited in only one valley of gapped graphene by the circularly polarized
pumping beam~\cite{Xiao,Mak2013}. In this case, the Hall exciton flows in a
monolayer or the Hall electron and hole currents in double layer,
are excited in only one valley of graphene. 
Therefore, the PME in a high strain induced pseudomagnetic field can
demonstrate the phenomena observed and studied for the charged particles in
high magnetic fields, namely, IQHE, FQHE~\cite{QHE}, and the state of
composite fermions at the filling factor $\nu = 1/2$~\cite{HLR,EM}.

In conclusion, we predict the zero magnetic field quantum Hall phenomena for
excitons in a high strain induced pseudomagnetic field. The Hall valley
flows of direct and indirect PME's, similar to Hall currents of charged
particles, can be excited in a mono or double layer of the gapped graphene,
respectively. In order to observe the quantum Hall effect for PMEs, one has
to measure the PME flows. We predicted the existence of PMEs and calculated
their properties. Moreover, we predicted the variety of collective
properties of PMEs analogous to 2DEG: IQHE, FQHE, and the state of composite
fermions at $\nu =1/2$. The spectrum of direct and indirect PMEs can be
studied by using the photoluminescence. One can use two optical methods to
observe the PME flows in mono and double layers: i. by measuring locally the
photoluminescence; ii. by measuring the shift of the angular distribution of
the photons emitted due to PME recombination as it was suggested for
polaritons in Ref.~\cite{BKL}. For a double layer with spatially separated
electrons and holes, one can measure the electron and hole currents in each
layer, forming a double layer. Electrons and holes forming indirect PMEs
contribute to these currents.

The observation of Landau level quantization and the gap $%
\Delta E$ in FQHE state for pseudomagnetoexcitons can be achieved by
two ways: i. using a THz spectroscopy for transitions between
neighboring Landau levels in the same band; ii. using an optical
spectroscopy for transitions between Landau levels of neighboring
bands with {$\left\vert n_{1}-n_{2}\right\vert =1,$ where $n_{1}$
and $n_{2}$} are quantum numbers of Landau levels in lower and upper
band, respectively. Fractional statistics of composite fermions can
be revealed in interference experiments.

The experimental observation of FQHE states for pseudomagnetoexcitons {is
possible} at temperatures{\ $T<\Delta E/k_{B}$,} where $k_{B}$ is the
Boltzmann constant and {$\Delta E$} is the energy between the excited and
ground states of the system. For the observation of FQHE for indirect
excitons {in suspended double layer graphene in pseudomagnetic field $%
B/e\sim 20\ \mathrm{T}$} at the lowest FQHE plateau (at Landau
filling $\nu =1/k$ with lowest $k$ and small interexciton
separations), where the interaction between indirect excitons are
almost the Coulomb one, plausible temperatures are the same as for
the FQHE for electrons, i.e. order of $20\ \mathrm{K.}$ {This is the
most optimal value for experiments. The plausible temperature} for
the double layer in the surrounding media with dielectric
susceptibility $\varepsilon _{d}$ is order of $20\
\mathrm{K}/\varepsilon _{d}$ (see Ref.~\cite{Bolotin} and references
therein). It is worth to notice that for the larger $k$ indirect
exciton interaction is weaker and corresponding {$\Delta E$} and
temperatures are lower and for direct pseudomagnetoexcitons even
more lower due to weaker exciton-exciton interactions.

The results of our study could provide a novel route to quantum Hall physics
of PMEs and PME-based valleytronics in graphene.  The predicted Quantum Hall
phenomena for the PMEs are  important, since they imply that the Quantum
Hall physics and valleytronics phenomena can be observed in the novel system
of neutral bosons without magnetic field.

The authors are grateful to G. Gumbs for the valuable  discussions. O.L.B.
and R.Ya.K. were supported by the U.S. ARO grant No. W911NF1810433 and PSC
CUNY under Grant No. 61538-00 49. The part of this work devoted to collective properties of pseudomagnetoexcitons  was supported by Grant No. 17-12-01393 of the Russian Science Foundation (Yu.E.L.). K.G.Z. is
grateful for support by a grant from the Julian Schwinger Foundation.


\begin{thebibliography}{99}
\bibitem{Geim_2010} F. Guinea, M.~I. Katsnelson, and A.~K. Geim, Nat.~Phys.
\textbf{6}, 30 (2010).

\bibitem{Guinea_Geim_2010} F. Guinea, A.~K. Geim, M.~I. Katsnelson, and
K.~S. Novoselov, \prb {\bf 81}, 035408 (2010).

\bibitem{Amorim} B. Amorim, A. Cortijo, F. de Juan, A.~G. Grushin, F.
Guinea, A. Guti\'{e}rrez-Rubio, H. Ochoa, V. Parente, R. Rold\'{a}n, P.
San-Jose, J. Schiefele, M. Sturla, and M.~A.~H. Vozmediano, Physics Reports
\textbf{617}, 1 (2016).

\bibitem{Vozmediano_2008} F. Guinea, M.~I. Katsnelson, and M.~A.~H.
Vozmediano, \prb {\bf 77}, 075422 (2008).

\bibitem{Jain2019} G.~A. Cs\'{a}thy and J.~K. Jain, Nat.~Phys. \textbf{15},
884 (2019).

\bibitem{Halperin2019} X. Liu, Z. Hao, K. Watanabe, T. Taniguchi, B.~I.
Halperin, and P. Kim, Nat.~Phys. \textbf{15}, 893 (2019).

\bibitem{Dean2019} J.~I.~A. Li, Q. Shi, Y. Zeng, K. Watanabe, T. Taniguchi,
J. Hone, and C.~R. Dean, Nat.~Phys. \textbf{15}, 898 (2019).

\bibitem{Guinea_Levy_2010} N. Levy, S. A. Burke, K. L. Meaker, M.
Panlasigui, A. Zettl, F. Guinea, A. H. Castro Neto, and M. F. Crommie,
Science \textbf{329}, 544 (2010).

\bibitem{Shalom} E. Sela, Ya. Bloch , F.~von Oppen, and M. Ben Shalom, \prl
{\bf 124}, 026602 (2020).

\bibitem{Lerner} I.~V. Lerner and Yu.~E. Lozovik, Sov.~Phys. JETP \textbf{51}%
, 588 (1980).

\bibitem{Ruvinsky} Yu.~E. Lozovik and A.~M. Ruvinsky, Phys. Lett. A \textbf{%
227}, 271 (1997); JETP \textbf{85}, 979 (1997).

\bibitem{another} Another method to obtain the homogeneous paseudomagnetic
field was proposed by using deformations with a triangular symmetry~[1].

\bibitem{Rytova} N.~S. Rytova, Proc. Moscow Stare University, Phys. Astron.
\textbf{3}, 30 (1967).

\bibitem{Keldysh} L.~V. Keldysh, JETP Lett. \textbf{29}, 658 (1979).

\bibitem{Landau} L.~D. Landau and E.~M. Lifshitz, \textit{Quantum Mechanics:
Non-Relativistic Theory} (Pergamon, Oxford, 1977).

\bibitem{Pedersen} T.~G. Pedersen, A.-P. Jauho, and K. Pedersen, \prb {\bf
79}, 113406 (2009).



\bibitem{GD} L.~P. Gorkov and I.~E. Dzyaloshinskii, Sov. Phys. JETP \textbf{%
26}, 449 (1967).

\bibitem{QHE} \textit{The Quantum Hall Effect}, edited by R.~E. Prange and
S.~M. Girvin (Springer-Verlag, New York, 1987).

\bibitem{Lukose} V. Lukose, R. Shankar, and G. Baskaran, \prl {\bf 98},
116802 (2007).


\bibitem{zhang89} S.~C. Zhang, T.~H. Hansson, and S. Kivelson, Phys. Rev.
Lett. \textbf{62}, 82 (1989).

\bibitem{fradkin_book} E. Fradkin, \textit{Field theories of Condensed
Matter Physics}, Cambridge University Press (Cambridge 2013).


\bibitem{Jain} J. Jain, \textit{Composite Fermions}, Cambridge University
Press (Cambridge 2007).

\bibitem{HLR} B.~I. Halperin, P.~A. Lee, and N. Read, \prb {\bf 47}, 7312
(1993).

\bibitem{EM} J.~P. Eisenstein and A.~H. MacDonald, Nature (London) \textbf{%
432}, 691 (2004).

\bibitem{Xiao} D. Xiao, W. Yao, and Q. Niu, \prl  {\bf 99}, 236809 (2007)

\bibitem{Mak2013} W. Yao, D. Xiao, and Q. Niu, \prb {\bf 77}, 235406 (2008).

\bibitem{BKL} O.~L. Berman, R.~Ya. Kezerashvili, and Yu.~E. Lozovik, \prb
{\bf 82}, 125307 (2010).

\bibitem{Bolotin} K.~I. Bolotin, F. Ghahari, M.~D. Shulman, H.~L. Stormer,
and P. Kim, Nature \textbf{462}, 196 (2009).
\end{thebibliography}
\end{document}


\title{Supplementary material: Strain induced quantum Hall effect of
excitons in graphene}
\author{Oleg L. Berman$^{1}$, Roman Ya. Kezerashvili$^{1}$, Yurii E. Lozovik$%
^{2}$, and Klaus G. Ziegler$^{3}$}
\affiliation{\mbox{$^{1}$Physics Department, New York City College
of Technology, The City University of New York,} \\
Brooklyn, NY 11201, USA \\
\mbox{$^{2}$Institute of Spectroscopy, Russian Academy of
Sciences,} \\
142190 Troitsk, Moscow, Russia \\
\mbox{$^{3}$Institut f\"ur Physik, Universit\"at Augsburg,\\
D-86135 Augsburg, Germany }}
\date{\today }
\maketitle

\section{Dirac particles}

\label{sect:Klaus}

In the presence of a strain field the Dirac equation was discussed in Ref.
\cite{amorim15}. When we apply this idea to independent particles we obtain
product states with single particle wave function $\psi_j(\mathbf{r}_j)$ of
energy $\mathcal{E}_j$
\begin{equation}
v_F\left(
\begin{array}{cc}
2 \Delta/v_{F} & i\hbar \partial _{x_j}+A_{x}(\mathbf{r}_j)+\hbar \partial
_{y_j}-iA_{y}(\mathbf{r}_j) \\
i\hbar \partial _{x_j}+A_{x}(\mathbf{r}_j)-\hbar \partial _{y_j}+iA_{y}(%
\mathbf{r}_j) & -2 \Delta/v_{F}%
\end{array}%
\right) \left(
\begin{array}{c}
\psi _{1,j}(\mathbf{r}_j) \\
\psi _{2,j}(\mathbf{r}_j)%
\end{array}%
\right) =\mathcal{E}_j\left(
\begin{array}{c}
\psi_{1,j}(\mathbf{r}_j) \\
\psi _{2,j}(\mathbf{r}_j)%
\end{array}%
\right) \ .
\end{equation}
In general, $\Delta$ can be caused not only by a strain field but also by
some other symmetry breaking source of the underlying lattice model.

The component $\psi_{1,j}$ of the Dirac spinor satisfies the Schr\"odinger
equation
\begin{equation}
\frac{1}{2m_j}[(i\hbar\partial_{x_j}+A_{x,j})^2+(i\hbar%
\partial_{y_j}+A_{y,j})^2]\psi_{1,j} =\frac{\mathcal{E}_j^{2}+\hbar
v_F^2B_{z}-4 \Delta^{2}}{2v_F^2m_j}\psi_{1,j}  \label{schroed0}
\end{equation}
and the second component is related to the first as
\begin{equation}
\psi_{2,j}(\mathbf{r}_j)=\frac{v_F}{\mathcal{E}_j+v_F\Delta}\left[i\hbar
\partial _{x_j}+A_{x}(\mathbf{r}_j)-\hbar \partial _{y_j}+iA_{y}(\mathbf{r}%
_j)\right] \psi_{1,j}(\mathbf{r}_j) \ .
\end{equation}
This implies that the eigenvalue of the Dirac equation reads $\mathcal{E}_j=%
\sqrt{2m_jv_F^2E_j-\hbar v_F^2B_{z}+ 4 \Delta^{2}}$, where $E_j$ is the
eigenvalue of the corresponding Schr\"odinger equation (\ref{schroed0}). In
the case of a homogeneous pseudomagnetic field we have $E_j=\hbar
(B_z/m_j)(n_j+1/2)$, which implies the relation
\begin{equation}
\mathcal{E}_{j,n_j}=\sqrt{2\hbar v_F^2B_zn_j+ 4 \Delta^{2}} \ .
\label{rel_dirac_schr}
\end{equation}

\section{Coordinate transformation}

\label{apa}

We introduce the vectors of the center-of-mass $\mathbf{R}$ and relative
motion $\mathbf{r}$ coordinates as
\begin{eqnarray}  \label{Rr}
\mathbf{R} = \frac{m_{1}\mathbf{r}_{1} + m_{2}\mathbf{r}_{2}}{m_{1} + m_{2}}
, \ \ \ \ \ \ \mathbf{r} = \mathbf{r}_{1} - \mathbf{r}_{2} .
\end{eqnarray}

Using these coordinates one can rewrite the operator of pseudomagnetic
momentum $\hat{\mathbf{P}}$ as
\begin{eqnarray}  \label{PRr}
\hat{\mathbf{P}} = -i \hbar \nabla_{\mathbf{R}} - \mathbf{B} \times \mathbf{R%
} - \frac{ \gamma \mathbf{B} \times \mathbf{r}}{2} ,
\end{eqnarray}
where
\begin{eqnarray}  \label{gam}
\gamma = \frac{m_{2} - m_{1}}{m_{2} + m_{1}} .
\end{eqnarray}
Since in the case of equal electron and hole effective masses at $m_{1} =
m_{2}$ one has $\gamma = 0$, the third term in the r.h.s. of Eq.~(\ref{PRr})
vanishes, and the gauge pseudomagnetic field acts only on the center-of-mass
of an electron and a hole and does not affect on their relative motion.

The operator $\hat{\mathbf{P}}^{2}$ is given by
\begin{eqnarray}  \label{PRr2}
\hat{\mathbf{P}}^{2} = - \hbar^{2} \nabla_{\mathbf{R}}^{2} + 2 i \hbar \left(%
\mathbf{B} \times \mathbf{R}\right)\cdot\nabla_{\mathbf{R}} + B^{2}R^{2} +
i\hbar \gamma \left(\mathbf{B} \times \mathbf{r}\right)\cdot\nabla_{\mathbf{R%
}} + \gamma \left(\mathbf{B} \times \mathbf{R}\right)\cdot \left(\mathbf{B}
\times \mathbf{r}\right) + \frac{ \gamma^{2}B^{2} r^{2}}{4} .
\end{eqnarray}

Using Eq.~(\ref{Rr}), the Hamiltonian (3) can be written in the following
form:
\begin{eqnarray}  \label{HamD}
\hat{H} = \hat{D}_{1} + \hat{D}_{2} + \hat{D}_{3} ,
\end{eqnarray}
where
\begin{eqnarray}  \label{D1}
\hat{D}_{1} = -\frac{\hbar^{2}\nabla_{\mathbf{R}}^{2}}{2M} - \frac{%
\hbar^{2}\nabla_{\mathbf{r}}^{2}}{2\mu} ,
\end{eqnarray}
\begin{eqnarray}  \label{D2}
\hat{D}_{2} = \frac{i \hbar}{2} \left[ \frac{2 \left(\mathbf{B}\times\mathbf{%
R}\right)\cdot\nabla_{\mathbf{R}}}{M} + \frac{(m_{2} - m_{1}) \left(\mathbf{B%
}\times\mathbf{R}\right)\cdot\nabla_{\mathbf{r}}}{m_{1}m_{2}} + \frac{\gamma
\left(\mathbf{B}\times\mathbf{r}\right)\cdot\nabla_{\mathbf{R}}}{M} + \frac{%
\left(m_{1}^{2} + m_{2}^{2}\right) \left(\mathbf{B}\times\mathbf{r}%
\right)\cdot\nabla_{\mathbf{r}}}{m_{1}m_{2}M} \right],
\end{eqnarray}
\begin{eqnarray}  \label{D3}
\hat{D}_{3} = \frac{B^{2}}{8}\left[\frac{R^{2}}{\mu} + \frac{%
2(m_{2}-m_{1})\left(\mathbf{r}\cdot\mathbf{R}\right)}{m_{1}m_{2}} + \frac{%
\left(m_{1}^{3} + m_{2}^{3}\right)r^{2}}{m_{1}m_{2}(m_{1}+m_{2})^{2}} \right]
.
\end{eqnarray}

and $M$ and $\mu$ are the total and reduced exciton masses, respectively,
given by
\begin{eqnarray}  \label{Mmu}
M = m_{1} + m_{2} , \ \ \ \ \ \ \frac{1}{\mu} = \frac{1}{m_{1}} + \frac{1}{%
m_{2}} .
\end{eqnarray}

\section{Derivation of the wavefunction and the energy of a PME}

\label{apaa}

Here we find the eigenfunctions and eigenvalues of the Hamiltonian $\hat{H}$%
~(3) 
without the assumption $m_{1} = m_{2}$. By introducing the coordinates for
the center-of-mass $\mathbf{R}$ and relative motion $\mathbf{r}$ of the
electron-hole system in the Hamiltonian (3) 
and expressing this Hamiltonian in terms of the operator $\hat{\mathbf{P}}$
after lengthly calculations presented in Supplementary Material S1
we obtain:
\begin{eqnarray}  \label{HamDP}
\hat{H} = \frac{1}{2M} \left[\hat{\mathbf{P}}^{2} - \frac{1}{m_{1}m_{2}}%
\left(\hbar M \nabla_{\mathbf{r}} - i \mathbf{S} (\mathbf{R},\mathbf{r})
\right)^{2}\right] ,
\end{eqnarray}
where
\begin{eqnarray}  \label{Sop}
\mathbf{S}(\mathbf{R},\mathbf{r}) = \frac{\mathbf{B}}{2}\times \left[%
(m_{2}-m_{1})\mathbf{R} + \frac{\left(m_{1}^{2} + m_{2}^{2}\right)\mathbf{r}%
}{M} \right] .
\end{eqnarray}

Since the Hamiltonian $\hat{H}$ commutes with the operator $\hat{\mathbf{P}}%
^{2}$, the eigenfunctions of $\hat{H}$ are also the eigenfunctions of $\hat{%
\mathbf{P}}^{2}$. While for an electron-hole pair in a magnetic field the
eigenfunctions of the Hamiltonian are the eigenfunctions of the
magnetoexciton momentum operator~\cite{GD,Lerner,Ruvinsky}, in the strain
induced pseudomagnetic field we present the eigenfunctions of $\hat{H}$ as
the eigenfunctions of $\hat{\mathbf{P}}^{2}$. Note that the $x$ and $y$
components of $\hat{\mathbf{P}}$ do not commute with one another.
Therefore, the eigenfunctions of $\hat{H}$ cannot be presented as the
eigenfunctions of $\hat{\mathbf{P}}$.

Let us find the eigenfunctions of the operator $\widetilde{H}_{0}$, defined
as
\begin{eqnarray}  \label{Ham0}
\widetilde{H}_{0} = \frac{\hat{\mathbf{P}}^{2}}{2M} = \frac{1}{2M} \left[-i
\hbar \nabla_{\mathbf{R}} - \mathbf{B} \times \mathbf{R} - \mathbf{A}_{0} %
\right]^{2} ,
\end{eqnarray}
where the vector $\mathbf{A}_{0}$ 
is given by
\begin{eqnarray}  \label{A0}
\mathbf{A}_{0} = \frac{ \gamma \mathbf{B} \times \mathbf{r}}{2} .
\end{eqnarray}
Let us find the eigenfunctions and eigenvalues of the Hamiltonian $%
\widetilde{H}_{0}$ for the both cases $\mathbf{A}_{0} = 0$ and $\mathbf{A}%
_{0} \neq 0$. If $\mathbf{A}_{0} = 0$, the eigenfunction of $\widetilde{H}%
_{0}$ is given by $\psi^{(0)} = \psi_{n,m}^{(0)}(\mathbf{R})$, which is the
wavefunction for a free particle of unit charge in the effective
pseudomagnetic field $2\mathbf{B}$ in the cylindrical gauge in
eigenvalue $E_{n}^{(0)}$ of $\widetilde{H}_{0}$ is defined as~\cite{Landau}
\begin{eqnarray}  \label{En0}
E_{n}^{(0)} = \frac{P_{n}^{2}}{2M} = \left(n + \frac{1}{2}\right) \hbar
\omega_{c} ,
\end{eqnarray}
where $\omega_{c} = 2 B/M$ is the cyclotron frequency for the motion of the
center-of-mass of a PME. The quantum numbers $n=0,1,2,\ldots $ and $%
m=0,1,2,\ldots$ for $\psi^{(0)} = \psi_{n,m}^{(0)}(\mathbf{R})$ and in Eq.~(%
\ref{En0}) are related to the motion of the center-of-mass of a PME.

If $\mathbf{A}_{0} \neq 0$, we define the scalar function $f(\mathbf{R})$ so
that $\mathbf{A}_{0} \equiv \nabla_{\mathbf{R}} f(\mathbf{R})$ and we have $%
f(\mathbf{R}) = \mathbf{A}_{0} \cdot \mathbf{R}$. In this case the
eigenvalue of $\widetilde{H}_{0}$ is the same as the eigenvalue at $\mathbf{A%
}_{0} = 0$ given by $E_{n} = E_{n}^{(0)} = P_{n}^{2}/(2M)$, and the
eigenfunction of $\widetilde{H}_{0}$ denoted as $\psi$ is given by
\begin{eqnarray}  \label{psi}
\psi \equiv \psi_{n,m} (\mathbf{R}) = \psi_{n,m}^{(0)}(\mathbf{R}) e^{if(%
\mathbf{R})/\hbar} .
\end{eqnarray}
We can see that
\begin{eqnarray}  \label{expf}
e^{if(\mathbf{R})/\hbar} = e^{i \mathbf{A}_{0} \cdot \mathbf{R}/\hbar} =
e^{i \gamma \left(\mathbf{B} \times \mathbf{r}\right)\cdot \mathbf{R}%
/2\hbar} = e^{i \gamma \left(\mathbf{B} \times \mathbf{R}\right)\cdot
\mathbf{r}/2\hbar} .
\end{eqnarray}
The eigenfunction $\Psi$ of the Hamiltonian $\hat{H}$ 
is given by
\begin{eqnarray}  \label{Psi}
\Psi = \psi_{n,m}^{(0)}(\mathbf{R}) e^{i \gamma \left(\mathbf{B} \times
\mathbf{R}\right)\cdot \mathbf{r}/2\hbar} \Phi (\mathbf{r}).
\end{eqnarray}
The function $\Phi (\mathbf{r})$ can be obtained from the solution of the
following equation:
\begin{eqnarray}  \label{HamDPeq}
\left[E_{n} - \frac{1}{2M m_{1}m_{2}}\left(\hbar M \nabla_{\mathbf{r}} - i
\mathbf{S} (\mathbf{R},\mathbf{r}) \right)^{2} + V(r) \right]e^{i \gamma
\left(\mathbf{B} \times \mathbf{R}\right)\cdot \mathbf{r}/2\hbar} \Phi (%
\mathbf{r}) = \mathcal{E} e^{i \gamma \left(\mathbf{B} \times \mathbf{R}%
\right)\cdot \mathbf{r}/2\hbar} \Phi (\mathbf{r}) ,
\end{eqnarray}
where $\mathcal{E} = E_{n} + \tilde{E}$ is the eigenvalue of the Hamiltonian
$\hat{H}$~(\ref{HamDP}). $\tilde{E}$ and $\Phi (\mathbf{r})$ can be obtained
from the solution of the following equation:
\begin{eqnarray}  \label{Phieq}
\left[- \frac{1}{2M m_{1}m_{2}}\left(\hbar M \nabla_{\mathbf{r}} - i \mathbf{%
S} (\mathbf{R},\mathbf{r}) \right)^{2} + V(r)\right] e^{i \gamma \left(%
\mathbf{B} \times \mathbf{R}\right)\cdot \mathbf{r}/2\hbar} \Phi (\mathbf{r}%
) = \tilde{E} e^{i \gamma \left(\mathbf{B} \times \mathbf{R}\right)\cdot
\mathbf{r}/2\hbar} \Phi (\mathbf{r}) .
\end{eqnarray}
Eq.~(\ref{Phieq}) can be rewritten as
\begin{eqnarray}  \label{Phieq11}
\left[\frac{1}{2\mu}\left(- i \hbar \nabla_{\mathbf{r}} - \frac{%
\left(m_{1}^{2} + m_{2}^{2}\right) \mathbf{B} \times \mathbf{r}}{2 M^{2}} -%
\frac{ \gamma \mathbf{B} \times \mathbf{R}}{2} \right)^{2} + V(r) \right]
\tilde{\Phi} (\mathbf{R},\mathbf{r}) = \tilde{E} \tilde{\Phi} (\mathbf{R},%
\mathbf{r}) ,
\end{eqnarray}
where $\tilde{\Phi} (\mathbf{R},\mathbf{r})$ is defined as
\begin{eqnarray}  \label{Phitild}
\tilde{\Phi} (\mathbf{R},\mathbf{r}) \equiv e^{i \gamma \left(\mathbf{B}
\times \mathbf{R}\right)\cdot \mathbf{r}/2\hbar} \Phi (\mathbf{r}) .
\end{eqnarray}
By applying to Eq.~(\ref{Phieq11}) the procedure similar to one was used to
find the eigenvalues and eigenfunctuions of the operator $\widetilde{H}_{0}$
given by Eq.~(\ref{Ham0}), we get
\begin{eqnarray}  \label{Phitild2}
\tilde{\Phi} (\mathbf{R},\mathbf{r}) = \tilde{\varphi}^{(0)}(\mathbf{r})
e^{i \gamma \left(\mathbf{B} \times \mathbf{r}\right)\cdot \mathbf{R}%
/2\hbar} .
\end{eqnarray}
Neglecting the electron-hole attraction 
one obtains $\tilde{\varphi}^{(0)}(\mathbf{r}) = \tilde{\varphi}_{\tilde{n},%
\tilde{m}}^{(0)}(\mathbf{r})$, where $\tilde{\varphi}_{\tilde{n},\tilde{m}%
}^{(0)}(\mathbf{r})$ is the wavefunction for a free particle of unit charge
the effective pseudomagnetic field $\mathbf{\tilde B} = \left(m_{1}^{2} +
m_{2}^{2}\right) \mathbf{B}/M^{2}$ in the cylindrical gauge in Refs.~\cite%
{Landau,Lerner,Ruvinsky}:
\begin{eqnarray}  \label{varphi}
\tilde{\varphi}_{\tilde{n},\tilde{m}}^{(0)}(\mathbf{r}) = \left[\frac{\tilde{%
n}!}{2 \pi \left(\tilde{n}+ |\tilde{m}|\right)!}\right]^{1/2} \frac{%
\exp\left(i\tilde{m}\phi \right)}{l}\left(\frac{r}{\sqrt{2}l}\right)^{|%
\tilde{m}|} L_{\tilde{n}}^{|\tilde{m}|}\left(\frac{r^{2}}{2l^{2}}\right)
\exp\left(- \frac{r^{2}}{4l^{2}}\right) \ ,
\end{eqnarray}
where $l = \sqrt{\hbar/{\tilde B}}$ is the pseudomagnetic length. In Eq.~(%
\ref{varphi}), $L_{\tilde{n}}^{|\tilde{m}|}$ denotes Laguerre polynomials.
The quantum numbers $\tilde{n} = \min(n_{1},n_{2})$, $\tilde{m} =
\left|n_{1} -n_{2}\right|$ are related to the relative motion of an electron
and a hole in the PME. The indexes $n_{1}$ and $n_{2}$ represent the
electron and hole quantum numbers, correspondingly. Let us mention that $l$
is measured in $\mathrm{m}$, since ${\tilde B}$ is measured in $\mathrm{kg/s}
$. Note that we consider a PME formed by an electron and a hole located in
the same type of valley, e.g., in the point K (or $K^{\prime}$) of the
Brillouin zone.

The value $\tilde{E}$ is the same as the energy of a free electron of mass $%
\mu$ in the effective pseudomagnetic field $\mathbf{\tilde B}$ in the
cylindrical gauge~\cite{Landau,Lerner}
\begin{eqnarray}  \label{En00}
\tilde{E}_{\tilde{n}} = \left(\tilde{n} + \frac{1}{2}\right) \hbar \tilde{%
\omega}_{c} ,
\end{eqnarray}
where $\tilde{\omega}_{c} = \left(m_{1}^{2} + m_{2}^{2}\right) \mathbf{B}%
/\left(M^{2}\mu\right)$ is the cyclotron frequency for the relative motion
of an electron and a hole in the PME.

Combining Eqs.~(\ref{Psi}),~(\ref{Phitild}), and~(\ref{Phitild2}), one can
see that the wavefunction of the electron-hole pair in the strain induced
gauge pseudomagnetic field, neglecting the electron-hole attraction, can be
written as
\begin{eqnarray}  \label{Psi111}
\Psi_{n,m,\tilde{n},\tilde{m}}(\mathbf{R},\mathbf{r}) = \psi_{n,m}^{(0)}(%
\mathbf{R})\tilde{\varphi}_{\tilde{n},\tilde{m}}^{(0)}(\mathbf{r}) e^{i
\gamma \left(\mathbf{B} \times \mathbf{r}\right)\cdot \mathbf{R}/2\hbar},
\end{eqnarray}
where $\gamma$ is defined by Eq.~(\ref{gam}), $\psi_{n,m}^{(0)}(\mathbf{R})$
is the wavefunction for a free particle in the effective pseudomagnetic
field $2\mathbf{B}$ in the cylindrical gauge in Refs.~\cite%
{Landau,Lerner,Ruvinsky}, $\tilde{\varphi}_{\tilde{n},\tilde{m}}^{(0)}(%
\mathbf{r})$ is the wavefunction for a free particle the effective
pseudomagnetic field $\mathbf{\tilde B} = \left(m_{1}^{2} + m_{2}^{2}\right)
\mathbf{B}/M^{2}$ in the cylindrical gauge in
Eq.~(\ref{varphi}).

In the expressions for $E_{n}$ and $\tilde{E}_{\tilde{n}}$ given by Eqs.~(%
\ref{En0}) and ~(\ref{En00}), respectively, $\omega_{c} = 2 B/M$ and $\tilde{%
\omega}_{c} = {\tilde B}/\mu$ are the cyclotron frequencies for the motion
of the center-of-mass and the relative motion of an electron and a hole in
the PME, respectively, and $n=0,1,2,\ldots$ and $\tilde{n}=0,1,2,\ldots$ are
the corresponding quantum numbers. In the case when $m_{1}=m_{2}\equiv m_{0}$
we have $\omega_{c}=\tilde{\omega}_{c}=B/m_{0}$. These expressions are used
to find the spectrum of the corresponding Dirac equation for the non
interacting electron-hole pair. Let us mention that for the PMEs in a high
strain-induced pseudomagnetic field we obtain that the energy spectrum of
both the motion of the center-of-mass and the relative motion of an electron
and a hole are quantized, in contrast to magnetoexcitons in a high magnetic
field, where the energy spectrum of the center-of-mass is continuous, and
the energy spectrum of the relative motion of an electron and a hole is
quantized \cite{GD,Lerner,Ruvinsky}.

In the case of the double layer when it is also possible that $m_{1}\neq
m_{2}$, Eq.~(9) from the main text can be written as:
\begin{eqnarray}  \label{difmas}
\mathcal{E}_{0n,\tilde{n}}=\sqrt{4\hbar v_{F}^{2}B_{z}n+4(\Delta _{1}+\Delta
_{2})^{2}}+ \sqrt{2\hbar v_{F}^{2}{\tilde{B}}_{z}{\tilde{n}}+ 4\left[ \Delta
_{1}\Delta _{2}/(\Delta _{1}+\Delta _{2})\right] ^{2}},
\end{eqnarray}
where $\Delta _{1}$ and $\Delta _{2}$ are the band gaps in the first and the
second graphene layers, respectively.
Therefore, Eq.~(\ref{difmas}) presents the quantized eigenenergy of the
non-interacting electron and hole in the strain induced pseudomagnetic field.

There are essential differences between the properties of a magnetoexciton
and a PME in a high magnetic and high strain-induced pseudomagnetic fields,
respectively. The Schr\"{o}dinger equation for a magnetoexciton in a
magnetic field $\mathbf{B}_{0}$ is are invariant with respect to the
translation and the gauge transformations~\cite{Ruvinsky}. This invariance
for a magnetic field results in the conservation of the operator of the
magnetic momentum of the magnetoexciton $\hat{\widetilde{\mathbf{P}}}%
=-i\hbar \nabla _{\mathbf{r}_{1}}-i\hbar \nabla _{\mathbf{r}_{2}}-\frac{e%
\mathbf{B}_{0}\times \left( \mathbf{r}_{1}-\mathbf{r}_{2}\right) }{2}$~\cite%
{GD,Lerner,Ruvinsky}. Since the operators $\hat{\widetilde{\mathbf{P}}}$ and
the Hamiltonian of a magnetoexciton commute, they have the same
eigenfunctions. If one acts by the Hamiltonian of the magnetoexciton on the
eigenfunction of $\hat{\widetilde{\mathbf{P}}}$ and employs the certain
variable change, the dependence of the resulting Hamiltonian on the
eigenvalue $\widetilde{\mathbf{P}}$ appears only in the term responsible for
the electron-hole Coulomb attraction as the replacement of $\mathbf{r}$ by $%
\mathbf{r} + \mathbf{r}_{0}$, where the continuously changing parameter $%
\mathbf{r}_{0}$ is directly proportional to the eigenvalue $\widetilde{%
\mathbf{P}}$, which can vary continuously from $0$ to infinity~\cite%
{Lerner,Ruvinsky}. Therefore, while the energy spectrum of a magnetoexciton
is discrete in the zeroth order with respect to the electron-hole
attraction, the energy spectrum of a magnetoexciton becomes a continuous
function of the eigenvalue $\widetilde{\mathbf{P}}$ in the first order
perturbation theory with respect to the electron-hole Coulomb attraction~%
\cite{Lerner,Ruvinsky}. The simultaneous invariance of the Schr\"{o}dinger
equation for a PME in the strain-induced pseudomagnetic field $\mathbf{B}$
with respect to the translation and the gauge transformations results in the
conservation of the operator of pseudomagnetic momentum $\hat{\mathbf{P}}%
=-i\hbar \nabla _{\mathbf{r}_{1}}-i\hbar \nabla _{\mathbf{r}_{2}}-\frac{%
\mathbf{B}\times \left( \mathbf{r}_{1}+\mathbf{r}_{2}\right) }{2}$. The
difference between the third terms of $\hat{\widetilde{\mathbf{P}}}$ and $%
\hat{\mathbf{P}}$ is caused by the fact that while the action of the
magnetic field on particles depends on the value and sign of charge of a
particle, the action of the strain-induced pseudomagnetic field on particles
does not depend on the value and sign of charge of a particle. Therefore,
the strain-induced pseudomagnetic field acts on an electron and a hole the
same way contrary to the magnetic field, which acts on an electron and a
hole differently. In this case, the resulting Hamiltonian does not
demonstrate the dependence on the continuously changing eigenvalue $\mathbf{P%
}$ only in the term responsible for the electron-hole Coulomb attraction as
the replacement of $\mathbf{r}$ by $\mathbf{r} + \mathbf{r}_{0}$, where the
continuously changing parameter $\mathbf{r}_{0}$ is directly proportional to
the eigenvalue $\mathbf{P}$. The strain-induced pseudomagnetic field acts on
a PME similar to the action of the magnetic field on two identical charged
particles. In the present Letter we demonstrated, that the latter leads to
the fact that the spectrum of a PME in a high strain-induced pseudomagnetic
field is discrete, in contrast to the spectrum of a magnetoexciton in a high
magnetic field which is continuous in the representation of magnetic
momentum~\cite{GD,Lerner}. Thus, Hall valley flows of direct and indirect
PMEs similar to Hall currents of charged particles can be excited in a mono
or double layer of the gapped graphene, respectively. These valley Hall
flows can be excited by circularly polarized light. Note that the Hall
valley flows of PMEs can be observed experimentally by studying the spatial
and angular characteristics of exciton photolumenescence. For spatially
indirect PMEs Hall flows can be measured in separated layers by analyzing
the electric currents of electrons and opposite currents of holes by
standard methodology.


\section{The energy of a PME}

\label{apb}

To find the energy of a direct and indirect exciton PME one should evaluate
the following matrix elements
\begin{eqnarray}
E_{0,0} &=&2\pi \int_{0}^{+\infty }\left[ \tilde{\varphi}_{0,0}^{(0)}(%
\mathbf{r})\right] ^{2}V(r)rdr,  \label{MatElements} \\
E_{0,1} &=&\int_{0}^{2\pi }d\phi \int_{0}^{+\infty }rdr\left[ \tilde{\varphi}%
_{0,1}^{(0)}(\mathbf{r})\right] ^{2}V(r), \\
E_{1,0} &=&2\pi \int_{0}^{+\infty }\left[ \tilde{\varphi}_{1,0}^{(0)}(%
\mathbf{r})\right] ^{2}V(r)rdr,
\end{eqnarray}%
where
\begin{eqnarray}
\tilde{\varphi}_{0,0}^{(0)}(\mathbf{r}) &=&\left[ \frac{1}{2\pi }\right]
^{1/2}\frac{1}{l}\exp \left( -\frac{r^{2}}{4l^{2}}\right) ,  \label{varphi11}
\\
\tilde{\varphi}_{0,1}^{(0)}(\mathbf{r}) &=&\left[ \frac{1}{2\pi }\right]
^{1/2}\frac{\exp \left( i\phi \right) }{l}\left( \frac{r}{\sqrt{2}l}\right)
\exp \left( -\frac{r^{2}}{4l^{2}}\right) , \\
\tilde{\varphi}_{1,0}^{(0)}(\mathbf{r}) &=&\left[ \frac{1}{2\pi }\right]
^{1/2}\frac{1}{l}\left( 1-\frac{r^{2}}{2l^{2}}\right) \exp \left( -\frac{%
r^{2}}{4l^{2}}\right) \
\end{eqnarray}
and $V(r)$ is the Coulomb or RK potential. In the case of an indirect PME in
the potential $V(r)$ the corresponding interparticle distance should be
replaced by the expression $\sqrt{r^{2}+D^{2}}$ \cite{BK2,BGK}, where $D$ is
the interlayer separation.

\section{The energy for direct PME's for the Coulomb and Rytova-Keldysh
potentials}

\label{apc}

The energy of a direct PME in a monolayer of gapped graphene double layer
can be calculated by substituting the Coulomb potential into Eq.~(6)
and one obtains
\begin{eqnarray}
E_{0,0}=-E_{0};\ \ E_{0,1}=-\frac{E_{0}}{2};\ \ E_{1,0}=-\frac{3E_{0}}{4}.
\label{Ecoulomb}
\end{eqnarray}%
%
%
In Eq.~(\ref{Ecoulomb}) $E_{0}$ is given by
\begin{eqnarray}
E_{0}=\frac{ke^{2}}{\varepsilon _{d}l}\sqrt{\frac{\pi }{2}},
\label{E0Coulomb}
\end{eqnarray}
where $l = \sqrt{\hbar/{\tilde B}}$ is the pseudomagnetic length.

The analytical expressions for the energy of a direct PME obtained using the
Rytova-Keldysh (RK) potential~\cite{Rytova,Keldysh} are the following:

\begin{eqnarray}
E_{0,0}=\frac{\pi ke^{2}}{\left( \varepsilon _{1}+\varepsilon _{2}\right)
\rho _{0}}\left[ e^{-\frac{l^{2}}{2\rho _{0}^{2}}}\text{Erfi}\left( \frac{l}{%
\sqrt{2}\rho _{0}}\right) -G\left( \left\{ \{0\},\{-\frac{1}{2}\}\right\}
,\left\{ \{0,0\},\{-\frac{1}{2}\}\right\} ;\frac{l^{2}}{2\rho _{0}^{2}}%
\right) \right] .  \label{E00}
\end{eqnarray}

\begin{eqnarray}
E_{0,1}=\frac{\pi ke^{2}}{\left( \varepsilon _{1}+\varepsilon _{2}\right)
\rho _{0}}\frac{1}{2\pi \rho _{0}^{2}}e^{-\frac{l^{2}}{2\rho _{0}^{2}}}\left[
-e^{\frac{l^{2}}{2\rho _{0}^{2}}}(2\rho _{0}^{2}-\sqrt{2\pi }\rho l)\text{ }%
-\pi (l^{2}-2\rho ^{2})\text{Erfi}\left( \frac{l}{\sqrt{2}\rho _{0}}\right)
+(l^{2}-2\rho _{0}^{2})\text{Ei}\left( \frac{l^{2}}{2\rho _{0}^{2}}\right) %
\right]  \label{E01}
\end{eqnarray}

\begin{eqnarray}
E_{1,0} &=&E_{0,0}-\frac{\pi ke^{2}}{\left( \varepsilon _{1}+\varepsilon
_{2}\right) \rho _{0}}\frac{1}{4\pi \rho _{0}^{4}}e^{-\frac{l^{2}}{2\rho
_{0}^{2}}}\left[ l^{4}(\gamma -1)+l^{3}\rho \sqrt{2\pi }e^{\frac{l^{2}}{%
2\rho _{0}^{2}}}-2l^{2}\rho ^{2}(1+\gamma )-e^{\frac{l^{2}}{2\rho _{0}^{2}}%
}(7\sqrt{2\pi }l\rho ^{3}-12\rho ^{4})\right.  \notag \\
&&-(6l^{2}\rho ^{2}-8\rho ^{4})\text{Ei}\left( \frac{l^{2}}{2\rho _{0}^{2}}%
\right) -\pi (l^{4}-8l^{2}\rho ^{2}+8\rho ^{4})\text{Erfi}\left( \frac{l}{%
\sqrt{2}\rho _{0}}\right) -(l^{4}+2l^{2}\rho ^{2})\text{ln}%
2+2(l^{4}-2l^{2}\rho ^{2})\text{ln}\frac{l}{\rho }  \notag \\
&&\left. +e^{\frac{l^{2}}{2\rho _{0}^{2}}}\left( \sqrt{2}l\rho _{0}-\rho
_{0}^{2}l^{2}{}_{1}F_{1}\left( 2,1;\frac{l^{2}}{2\rho _{0}^{2}}\right)
\right) \right] ,  \label{E10}
\end{eqnarray}%
where $\gamma $ is Euler constant, Erfi$\left( x\right) $ is the imaginary
error function, the Maijer $G-$function, Ei$(x)$ is the exponential integral
function, and ${}_{1}F_{1}\left( a,b;x\right) $ is the Kummer confluent
hypergeometric function.

\section{The energy for indirect PME's for the Coulomb potential}

\label{apd}

\begin{eqnarray}  \label{Enmagin}
E_{0,0}(D) &=&-E_{0}\exp \left[ \frac{D^{2}}{2l^{2}}\right] \mathrm{Erfc}%
\left[ \frac{D}{\sqrt{2}l}\right] ,  \label{m2} \\
E_{0,1}(D) &=&-E_{0}\left[ \left( \frac{1}{2}-\frac{D^{2}}{2l^{2}}\right)
\exp \left[ \frac{D^{2}}{2l^{2}}\right] \mathrm{Erfc}\left[ \frac{D}{\sqrt{2}%
l}\right] +\frac{D}{\sqrt{2\pi }l}\right] , \\
E_{1,0}(D) &=&-E_{0}\left[ \left( \frac{3}{4}+\frac{D^{2}}{2l^{2}}+\frac{%
D^{4}}{4l^{4}}\right) \exp \left[ \frac{D^{2}}{2l^{2}}\right] \mathrm{Erfc}%
\left[ \frac{D}{\sqrt{2}l}\right] -\frac{D}{2\sqrt{2\pi }l}-\left( \frac{D}{%
\sqrt{2}l}\right) ^{3}\frac{1}{\sqrt{\pi }}\right] ,
\end{eqnarray}%
where Erfc$(x)$ is the complementary error function and $E_{0}$ is given by (%
\ref{E0Coulomb}). These expressions partially concise with the expressions
obtained in the case of uniform magnetic field~\cite{Ruvinsky}.

In Table ~(\ref{tab2}) are given results of calculations for the energies
using Eqs.~(\ref{Ecoulomb}) - (S43).
\begin{table}[t]
\caption{Calculations performed for the gapped graphene, $\Delta =0.25$ eV, $%
\protect\varepsilon =13$ and the value of magnetic length $l$\ that
corresponds to $B/e =50$ T. Two gapped graphene layers are separated by $D =
1.7$ nm.}
\label{tab2}
\begin{center}
\begin{tabular}{ccccc}
\hline\hline
Energy & Potential & Monolayer & 2 Layers & Landau Level, eV \\ \hline
$E_{0,0},$ meV & RK & $27.001$ & $21.185$ & 1.25 \\
& Coulomb & $27.097$ & $21.187$ &  \\
$E_{0,1},$ meV & RK & $13.548$ & $13.013$ & 1.30 \\
& Coulomb & $13.550$ & $13.014$ &  \\
$E_{1,0},$ meV & RK & $20.228$ & $15.128$ & 1.32 \\
& Coulomb & $20.322$ & $15.130$ &  \\ \hline\hline
\end{tabular}%
\end{center}
\end{table}

\section{The role of spins}

\label{ape}

 Since the spin-orbit interaction in graphene is
negligible, in contrast to TMDC, the degeneracy does not lead to the
splitting of the corresponding levels. Therefore, the problem of a
pseudomagnetoexciton can be solved with the electron spin up and
hole spin down in the upper and lower bands of gapped graphene,
respectively. The opposite spins of the electron and hole is due to
the conservation of the electron spin projection in an allowed
optical transition. Therefore, we are considering
pseudomagnetoexcitons in the singlet state. The valley is fixed due
to the excitation by a laser beam with circular polarization. It is
worth mentioning that the fixation of the valley leads to the
fixation of the spin.

\section{The stability of quantum Hall effects for PME}

\label{apf}

The lifetime of spatially indirect excitons increases with the
number of intermediate insulating layers between two graphene layers
where electron and hole are located due to the decrease in the
overlap of the wave functions. For example, for 3 intermediate
insulating layers this time is already 10 nanoseconds. In comparison
with direct excitons it is by 4 orders of magnitude larger, and each
layer increases the lifetime by about 1.3 orders of
magnitude~\cite{Calman}. The lifetime of direct excitons is
essentially shorter but earlier it was proposed the method to
essentially increase the lifetime of direct excitons by employing
photonic engineering, namely by location of the 2D layer in
subwavelength optical cavity~\cite{Voronova}.

The characteristic time necessary for the formation of the FQHE state can be
estimated as $t \sim \hbar/\Delta E$, where $\Delta E$ is characteristic
energy defined the stability of FQHE (see below) - the energy difference
between the state with one hole excitation and the ground state of the
system (described by the Laughlin-type wave function). The estimate of $%
\Delta E$ below shows that this time is essentially smaller than the
lifetime of the exciton. This allows the observation of FQHE for
pseudomagnetoexcitons. The allowed temperatures for the observation
of FQHE are $T < \Delta E/k_{B}$.

Note that there are different estimates of the gap $\Delta E$
associated with different excitations over the ground state of the
system in  the FQHE state- i. the creation of a composite hole; ii.
the creation of a Coulomb interacting \emph{e-h} pair; iii. the
creation of a collective excitation of the pseudoskyrmions type. The
estimates for different excitations do not coincide, but they are of
the same order of magnitude! (see Ref.~\cite{Toke} and references
therein).

The stability of the integer quantum Hall effect for
pseudomagnetoexcitons is determined by the energy gap between Landau
levels, which is proportional to the pseudomagnetic field. Landau
level quantization for pseudomagnetoexcitons can be revealed by
optical spectroscopy. To reveal IQHE Hall quantization by transport
experiments for pseudomagnetoexcitons the following are needed: i.
the plateau formation due to localization on impurities; ii. all
states at Landau level must be filled. The latter is impossible for
Bose quasiparticles, but for pseudomagnetoexcitons it is possible
only for composite fermions formed due to exciton-exciton
interactions. The composite fermion stability is also defined by the
energy gap $\Delta E$ proportional to the characteristic energy of
interaction of excitons $\Delta E$ at a distance $r=l$ corresponding
to the filling of the Landau level $\nu =1$. Thus the energy
corresponding to forming IQHE plateau in contrast to fermion,
electron system has the order of $\Delta E\sim e^{2}/\varepsilon
_{d}l$ and the necessary temperatures  for the observation IQHE for
pseudomagnetoexcitons in transport experiments are $T < \Delta
E/k_{B}$.